\documentclass[manuscript]{emulateapj}
\usepackage{graphicx}

\shorttitle{Evolution of galaxy groups in the Illustris simulation}
\shortauthors{M. Raouf et al.}

\begin{document}


\title{Evolution of galaxy groups in the Illustris simulation}


\author{Mojtaba Raouf\altaffilmark{*} and Habib G. Khosroshahi\altaffilmark{}}
\affil{School of Astronomy, Institute for Research in Fundamental Sciences (IPM),
	Tehran, 19395-5746, Iran}
\and
\author{A.~Dariush\altaffilmark{}}
\affil{Institute of Astronomy, University of Cambridge, Madingley Road, Cambridge CB3 0HA, UK}
\email{*m.raouf@ipm.ir}




\begin{abstract}
	
	We present the first study of evolution of galaxy groups in the Illustris simulation. We focus on dynamically relaxed and unrelaxed galaxy groups representing dynamically evolved and evolving galaxy systems, respectively. The evolutionary state of a group is probed from its luminosity gap and separation between the brightest group galaxy and the center of mass of the group members. We find that the Illustris simulation, over-produces large luminosity gap galaxy systems, known as fossil systems, in comparison to observations and the probed semi-analytical predictions. However, this simulation is equally successful in recovering the correlation between luminosity gap and luminosity centroid offset, in comparison to the probed semi-analytic model. We find evolutionary tracks based on luminosity gap which indicate that a large luminosity gap group is rooted in a small luminosity gap group, regardless of the position of the brightest group galaxy within the halo. This simulation helps, for the first time, to explore the black hole mass and its accretion rate in galaxy groups. For a given stellar mass of the brightest group galaxies, the black hole mass is larger in dynamically relaxed groups with a lower rate of mass accretion. We find this consistent with the latest observational studies of the radio activities in the brightest group galaxies in fossil groups. We also find that the IGM in dynamically evolved groups is hotter for a given halo mass than that in evolving groups, again consistent with earlier observational studies. 
	
\end{abstract}

\section{Introduction} 

In earlier contributions, we have highlighted the importance of identifying fossils groups or dynamically relaxed groups and clusters \citep{Khosroshahi2006,Khosroshahi2007,Khosroshahi2014,Raouf2014,Miraghaei2014,Gozaliasl2014,Khosroshahi2016}. Adapting these classifications and continued studies enable us (I) to explore if these systems truly follow a different evolutionary path in their galaxy or halo properties, (II) to employ these galaxy systems and their statistical properties to identify the best possible model of galaxy formation and evolution, generally implemented in cosmological simulations and (III) to better understand the galaxy-halo connection. 

Dark matter simulations have shown that galaxies in compact groups should merge into a single massive galaxy within a Gyr \citep{Barnes1989, Bode1993}. Consequently, an elliptical galaxy is formed, developing a large luminosity gap while the X-ray emitting halo remains unaffected by merging \citep{Ponman1994}. Such groups are known as fossil groups in which the essential observational tracers have been identified including the luminosity gap between the first and second brightest galaxy group members and the presence of an extended, i.e. group scale, X-ray emission with a luminosity of at least $L_{X,bol}\approx 10^{42} $~h$_{50}^{-2}$ ~erg~s$^{-1}$ \citep{Jones2003}. There are several studies in the literature focusing on the detailed characterization and properties of fossil groups base on X-ray and optical observations \citep{Khosroshahi2004,Sun2004,Ulmer2005,Khosroshahi2006,Khosroshahi2007,Miraghaei2014}, cosmological simulations \citep{Yoshioka2004,Milosavljevic2006,VandenBosch2007,VonBenda-Beckmann2008,Deason2013}, semi-analytical models \citep{Sales2007,Dariush2007,Diaz2008,Dariush2010,Raouf2014} and hydrodynamical simulations \citep{DOnghia2005,Cui2011}. Recent study of \cite{Khosroshahi2014} reveals that a diffuse extended X-ray emission beyond the optical size of the brightest group galaxy, exists specially when a large magnitude gap is present.

\citet{Khosroshahi2006} presented evidences that the majority of Brightest Group Galaxy (BGG) dominating fossil galaxy groups have non-boxy isophotes which could point to wet, or gas rich, nature of galaxy merger in their evolutionary history. \cite{Smith2010} employed a large sample of BGGs observed with the Hubble Space Telescope and found the trend in the luminosity gap (as an indication for the dynamical age of the system) and the isophotal shape of the BGGs, to be consistent with earlier study of \citet{Khosroshahi2006}. Furthermore, in comparison with the general population of galaxy groups, \citet{Khosroshahi2007} show that for a given optical luminosity, fossil groups not only contain hotter Intra Galactic Medium(IGM) for a given halo mass, but also their dark matter halo is more concentrated, all pointing at their relatively earlier formation epoch. In addition, the study of scaling laws in fossil groups indicate that they mostly follow the trend of galaxy clusters which is likely to be driven by dynamically relaxed state of cluster core. It worths highlighting an apparent conflict, as \citet{Voevodkin2010} show that there is no noticeable difference between the X-ray luminosity of the fossils and non-fossils for a given optical luminosity. While more recent studies support the latter \citep{Aguerri2011,Proctor2011,Harrison2012,Girardi2014}, however, the apparent contradiction could primarily be originated from the sample selection, and is due to the fundamental differences between galaxy groups, which forms the basis for \citet{Khosroshahi2007}, and galaxy cluster sample by \cite{Santos2007} which forms the basis for contrasting studies. In a Lambda-CDM model galaxy clusters are generally young assembly of galaxies while galaxy groups can be old and young depending on whether they survive major mergers during the hierarchical cosmic evolution.  

\begin{figure}
	\includegraphics[width=0.47\textwidth]{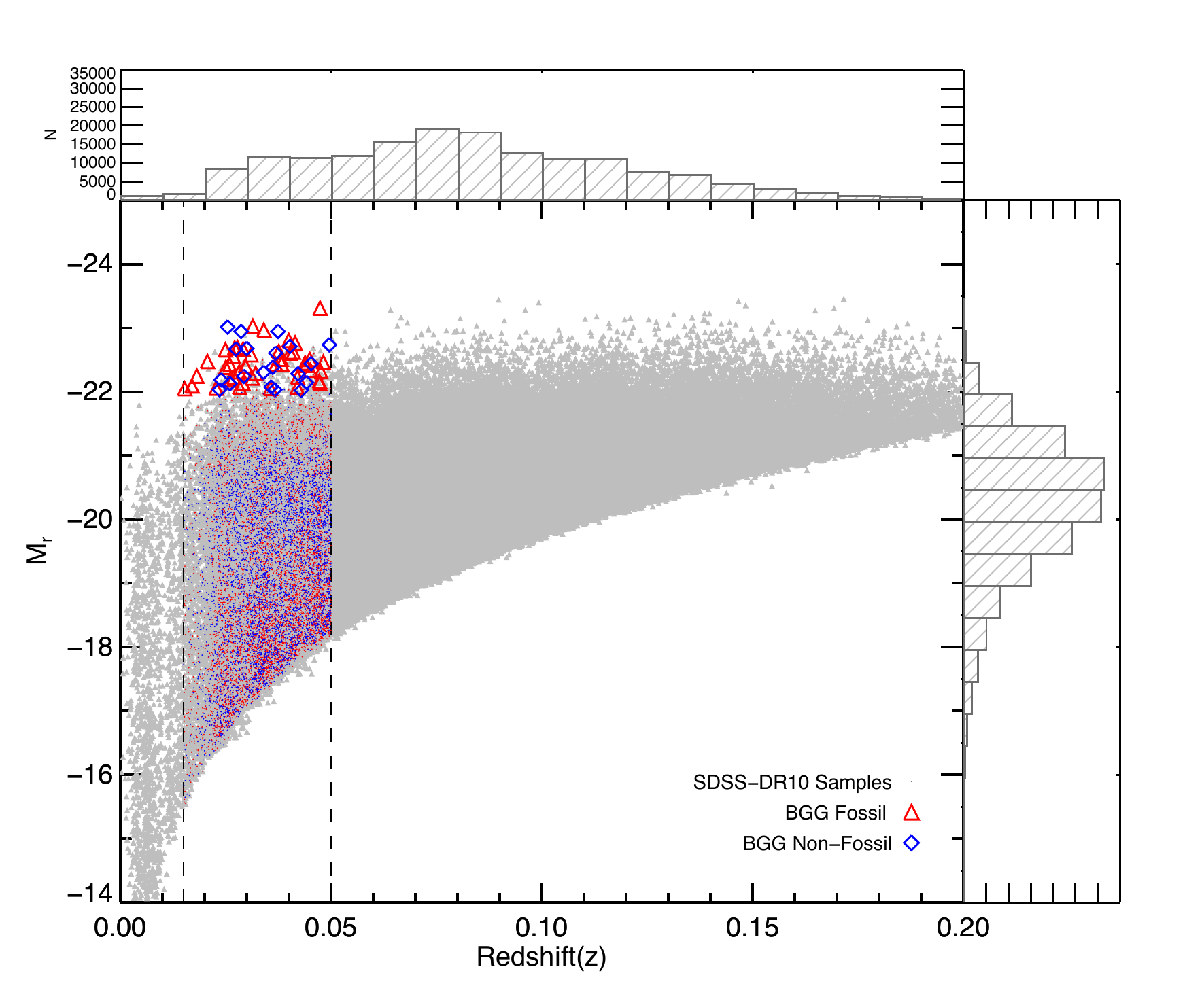}
	\caption{Absolute r-band magnitude distribution as a function of redshift based on SDSS DR10 group catalogue of \citet{Tempel2014}. The red triangles show the BGGs belonging to fossil groups ($\Delta M_{12}\gtrsim2$) while blue diamonds  show non-fossil ($\Delta M_{12}<0.5$) galaxy groups. In addition, The red and blue dots represent the members of fossil and non-fossil galaxy groups, respectively.}
	\label{SDSS:fig}
\end{figure} 

In recent years, cosmological simulations have offered the necessary tools to address open questions, regarding the formation and evolution of galaxies. This is generally achieved through Semi-Analytical Models (SAMs) and hydrodynamical models. In semi-analytic approach the baryonic matter properties are calculated on the basis of analytical prescription in a post-processing procedure built on the merger tree \citep{Croton2006,Bower2006,DeLucia2007,Guo11}. In hydrodynamical approach, baryons directly interact and co-evolve with the dark matter particles within the cosmological volume. Although, the hydrodynamic approach has the upper hand in dealing with baryonic matter that can be directly linked to the gas properties ( such as cooling , heating and feedback process in and around galaxy halos \citep[and their references]{Springel2005a,Vogelsberger2014a}, but the semi-analytic approach is computationally inexpensive compared to the hydrodynamic and facilitate to construct sample of galaxies which are an order of magnitude larger than the same allowed by hydrodynamical simulations. Furthermore, the SAMs are more suitable for adding in new physics and assessing the impact. 

A number of authors have suggested ways in which, radio-AGN heating is powerful enough to expel a fraction of baryons from the galaxy groups or clusters  \citep{Croton2006,Bower2008}. Observationally, some studies show that radio-AGN heating could account for the missing baryons in galaxy groups \citep{Oklopcic2010,Giodini2010}.  A useful approach for understanding the role of AGN feedback in galaxy evolution is to connect the astrophysical parameters related to the AGN feedback to observable quantities and make predictions which can be verified by the existing or future observations. 

In a recent study \citep{Raouf2014}, we established a set of four observationally measurable parameters using the semi-analytic models of \cite{Guo11}, based on the Millennium Simulation, which can be used in combination, to identify a subset of galaxy groups which are dynamically old, with a very high statistical probability. We argued that a sample of fossil groups selected based on luminosity gap will result in a contaminated sample of old galaxy groups. However, by adding constraints on the offset between the group luminosity centroid and the BGG position, we considerably improved the age-dating method for galaxy groups and clusters, in comparison to the method based on the luminosity gap only.

\begin{figure}
	\includegraphics[width=0.5\textwidth]{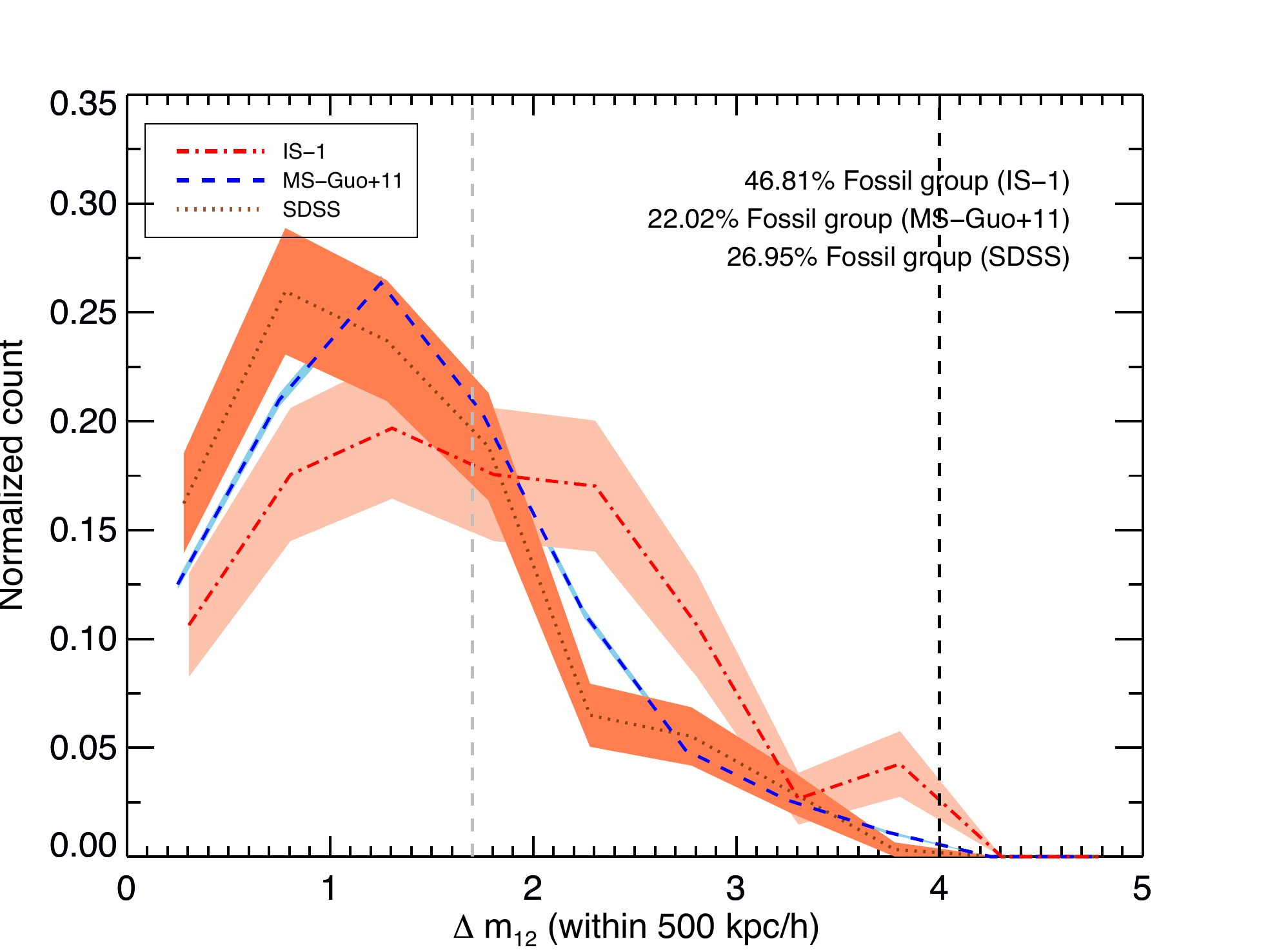}
	\caption{Distribution of luminosity gap ($\Delta m_{12}$) for 3 categories; SDSS (brown dotted line), MS-Guo+11 (blue dashed line) and IS-1 (red dotted-dashed line). Colour filled regions represent the poison errors for each bin in the three categories. The IS-1 distribution shows that a large fraction of galaxy groups have large luminosity gap in contrast of the other two. The fraction of fossil groups ($\Delta m_{12} > 1.7 \ mag$) in  IS-1, MS-Guo+11 and SDSS are $\approx$47, 22 and 27 per cent, respectively. The vertical gray dashed line marks the luminosity gap for fossil galaxy groups and the black dashed line marks the completeness limit of the luminosity gap for the  SDSS sample. }
	\label{gap12}
\end{figure} 

\begin{figure}
	\includegraphics[width=0.47\textwidth]{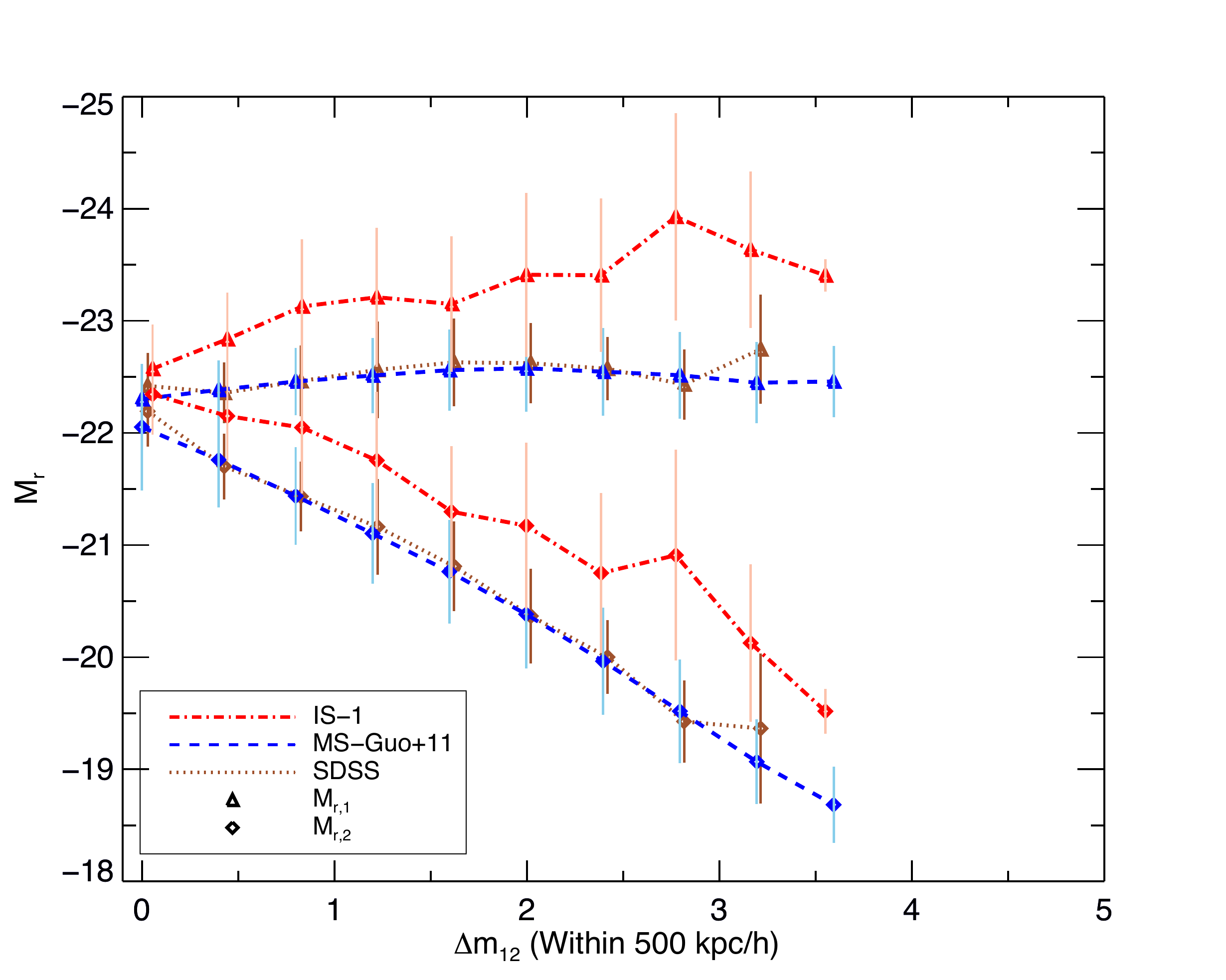}
	\caption{Absolute r-band magnitude of the first (triangle) and the second ranked (diamond) galaxies as function of luminosity gap, ($\Delta m_{12}$), for 3 categories of SDSS (brown dotted line), MS-Guo+11 (blue dashed-line) and IS-1 (red dotted-dashed-line). The data points illustrate the average bin of data and the error bars present the standard deviation of bind  data points in each categories.}
	\label{BF:fig}
\end{figure} 

In this study, the main focus is to explore the Illustris in the context of luminosity gap formation and the advantage that this simulation may offer in providing a hydro based measure of accretion rate and super massive black hole mass.  Thus we prepare an observed sample of galaxy systems using the Sloan Digital Sky Survey (SDSS) data release 10 \citep{Ahn2014} and similarly in the Millennium Simulation \citep{Springel2005} joined with \cite{Guo11} Semi-Analytic model and high resolution gravitational and hydrodynamical simulation of Illustris-1 \citep{Vogelsberger2014a}. Section 2 describes the data that we are using for simulation and observation. In Section 3 we describe our analysis and compare the luminosity gap of two brightest group galaxies and the map of luminosity gap -- centroid off-set. Finally in Section 4, we present summery of our results and conclusion. In this paper a $\Lambda CDM$ cosmology with $\Omega_m = 0.3$, $\Omega_{\Lambda} = 0.7$ and $H_0 = 100 h$ km $s^{-1}$ $Mpc^{-1}$ where $h=0.7$ is assumed.

\section{Data \& Simulations}

\subsection{Observations: SDSS group catalogues}

We use the legacy archive of the Sloan Digital Sky Survey-III, Data Release 10 \citep[SDSSIII-DR10] {Ahn2014} which covers 14,555 $deg^2$ in imaging data containing 469053874 unique objects for which 3276914 spectra were measured. 
In this study, we use the FOF group/cluster catalogue of \citet{Tempel2014} for redshift range between 0.015 and 0.05 (Figure \ref{SDSS:fig}). Morphologically, all BGGs in our sample are elliptical with absolute r-band magnitude of $M_r(BGG)< -22$ mag and reside in halos with masses  (referred in the catalogue as "massNFW") $\geq 10^{13} M_{\odot}$. Groups contain at least 4 spectroscopic members. With these constraints the observational sample contain 300 galaxy groups/clusters.
We estimate the offset ($D_{off}$) between the BGG location and the luminosity centroid, using the r-band magnitude of group's spectroscopic members and their coordinates. The luminosity gap in the r-band is obtained within 500 kpc/h radius from the BGG. Figure \ref{SDSS:fig} shows the distribution of absolute magnitude in the r-band vs. the redshift for all galaxies (gray points) as well as the selected sample in the SDSS (blue and red points). 

\subsection{Simulations: Illustris-1 and Millennium Simulations}

We use the public release of the Illustris-1 Simulation \citep[Hereafter: IS-1]{Vogelsberger2014a}
a series of gravity as well as hydrodynamics realizations of a $(106.5$ $Mpc)^3$ cosmological volume that contains $1820^{3}$ gas cell and $1820^{3}$ dark matter particles, run with the AREPO code \citep{Springel2010}.
The highest-resolution run of the Illustris-1 handles the dark matter (DM) component with a mass resolution of $m_{DM} = 6.3 \times 10^6M_{\odot}$ and a baryonic component with $m_{baryon} = 1.6 \times 10^6M_{\odot}$ in 136 snapshots from z = 127 to z = 0 by adopting cosmological parameters consistent with the latest Wilkinson Microwave Anisotropy Probe 9 observations \citep[WMAP-9]{Hinshaw2013}. Halos, subhalos, and their basic properties have been identified with the FOF and SUBFIND algorithms \citep{Davis1985,Springel2001,Dolag2009} at every stored snapshots. Based on halo mass limit of $M_{halo}\geq 10^{13}-10^{14} M_{\odot}$ on galaxy systems with BGG absolute r-band magnitude of $M_r(BGG)< -22$ mag and multiplicity of at least 4 members for groups, the number of galaxy group in the present epoch is reduced to $\sim$ 190 systems containing $\sim$ 15000 galaxies. See \cite{Nelson2015} for more detailed description of the galaxy group catalogue properties. 

In addition, we are using the Millennium Simulation \citep[MS]{Springel2005} joined with the \cite{Guo11} Semi-Analytical Model (Hereafter: MS-Guo+11) to extract galaxy properties. The cosmological model adopted in the Millennium Simulation is consistent with the first Wilkinson Microwave Anisotropy Probe 1 data \citep[WMAP-1]{Spergel2003} (note that the value of ${\sigma}_8$ is assumed to be greater than its present value of 0.82 given by WMAP-9 that is not strongly affects in this study). The simulation box $(500 h^{-1} Mpc)^{3}$ contains $2160^{3}$ particles and presents the mass resolution of $8.6 \times 10^8 h^{-1}M_{\odot}$. The dark matter merger trees within each simulation snapshot (64 snapshots in total) are spanned approximately logarithmically in time between $z=127$ and $z=0$ and extracted from the simulation using combination of FoF \citep{Davis1985} and SUBFIND \citep{Springel2001} halo finders algorithms. The gas and stellar components of galaxies in dark matter halos are constructed semi-analytically, based on laying a series of couples differential equations on top of the halo merger trees. In this study, we use  \citet{Guo11} semi-analytical model at the present epoch which contains  $\sim 23000$ galaxy groups/clusters with at least 4 members and halo mass above 10$^{13}$~M$_{\odot}$ to $\sim 10^{14} M_{\odot}$ with BGG absolute r-band magnitude of  $M_r(BGG)< -22$ mag and  $\sim2$ million galaxies.

Luminosity centroid for the simulations is defined base on  $X_L= \sum X_i L_i/ \sum L_i$, where $L_i$ is the luminosity of a galaxy within a group in the r-band and $X_i$ is the projected coordinate of each galaxy within the radius of  $r_{200}$.
Finally, we use the  r-band magnitude of the group members and their coordinates to obtain the luminosity gaps within 500 kpc/h radius in each simulation.

\section{Result}

\subsection{The Luminosity Gap} \label{gap:sec}

In the previous studies of fossil groups, the luminosity gap between the two most luminous galaxies, located within a given physical radius of the group center (e.g. 0.5 R$_{200}$), has been used as a statistical tool to probe the accuracy of a number of semi-analytic galaxy formation models in cosmological simulations \citep{Dariush2007,Dariush2010,Smith2010,Raouf2014,Gozaliasl2014}.  \cite{Dariush2007} used \cite{Croton2006} SAM in MS studies, to predict that fossil systems could be found in significant numbers (3--4 per cent of the population) even in quite rich clusters. Other probes have also been proposed, for instance \cite{Dariush2010} introduced $\Delta$m$_{14} \geq 2.5$, i.e. the luminosity gap between the first and fourth brightest galaxies within $0.5R_{200}$, as opposed to the conventional $\Delta$m$_{12} \geq 2.0$. 
\cite{Smith2010} combined a series of observational data to study the luminosity gap statistics within a radius of $\sim$640 kpc a sample of 59, intermediate mass, galaxy clusters. They show that base on the  luminosity gap parameters, 8$\pm3$ per cent of the sample are fossil systems. Recently, \cite{Gozaliasl2014} studied luminosity gap distribution using a large sample of X-ray galaxy groups (129 groups) spanned over redshift $z\le 1$ in the  XMM-LSS X-ray observations and the CFHT follow-up optical observations. They found that 22$\pm6$ per cent of groups at z $\leq$ 0.6 are fossils.
\begin{figure}
	\includegraphics[width=0.5\textwidth]{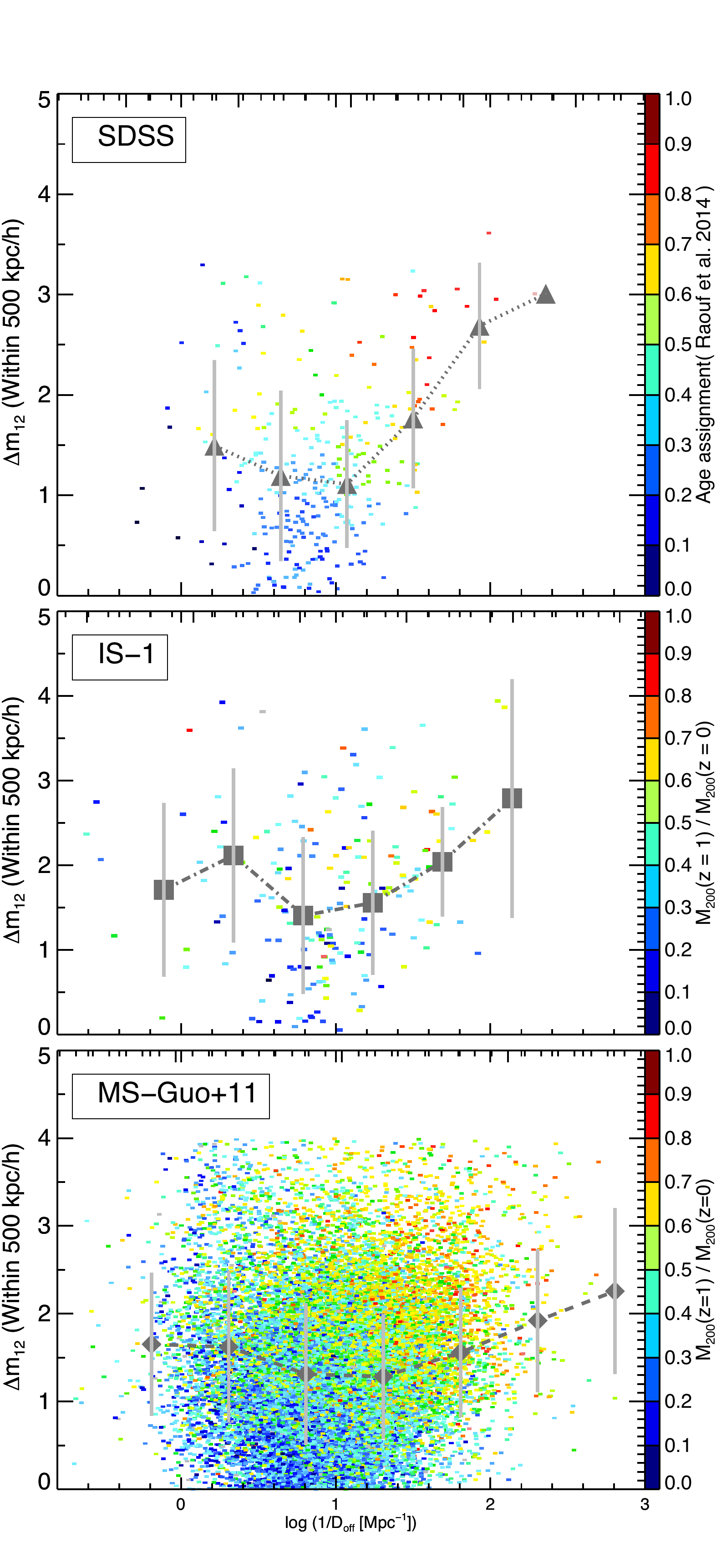}
	\caption{Distribution of the luminosity gap parameter as function of inverse centroid offset ($1/D_{off}$) for the SDSS, IS-1 and MS-Guo+11 samples. The colour coding is base on the halo age. For the SDSS sample this is based on the method of age assignment in \citep{Raouf2014}. In the simulations, this is based on the fraction of halo mass in z=1 to the final mass (z=0). Thus, the red colour represent the oldest galaxy groups and the blue marks the youngest galaxy groups. The gray dotted, dot-dashed and dashed lines show the mean of the binned data points with the standard deviation error bar ($\sigma$) for SDSS, IS-1 and MS-Guo+11, respectively. }
	\label{gap_off}
\end{figure}

\begin{figure*}
	\centering
	\includegraphics[width=0.8\textwidth]{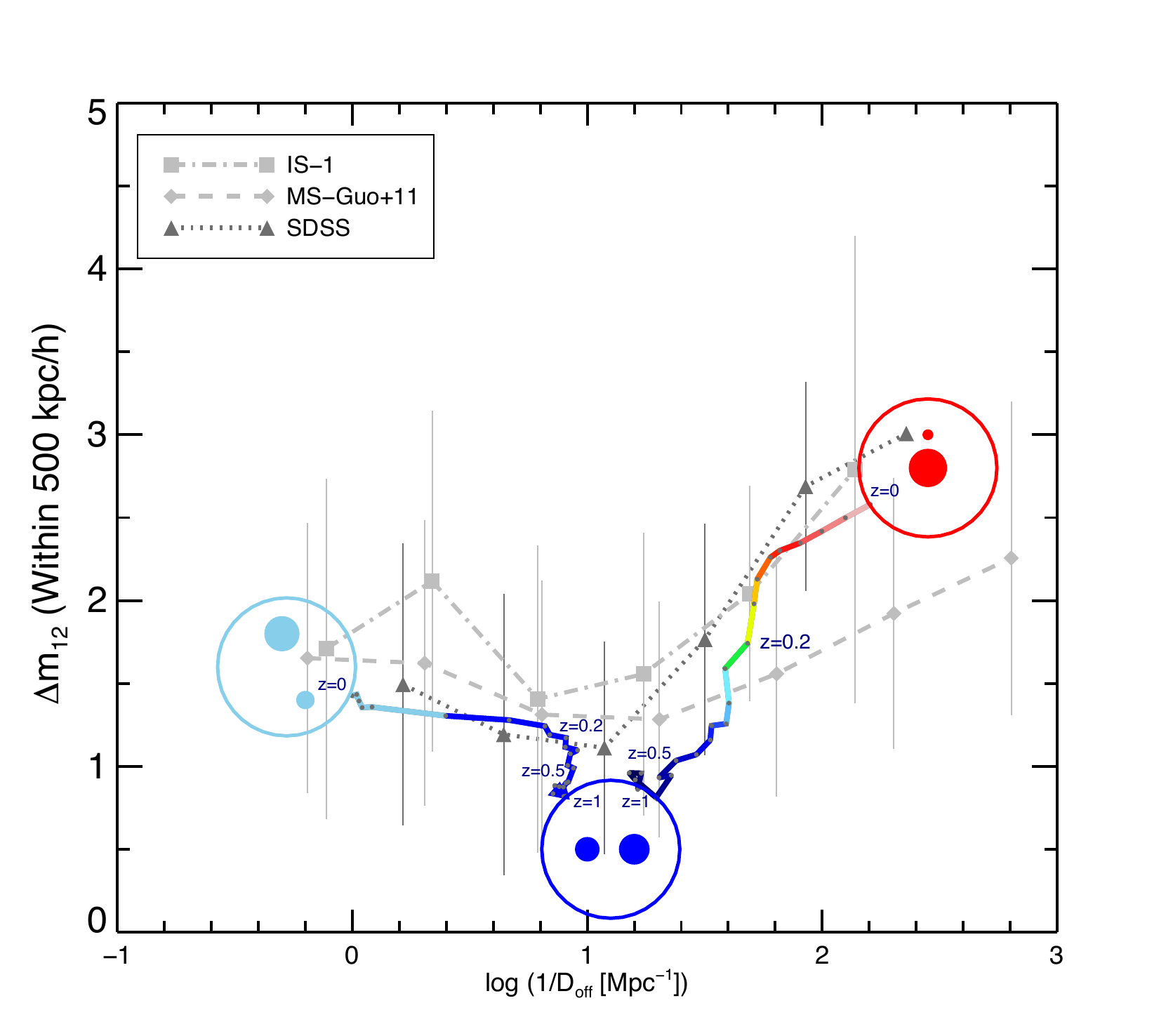}
	\caption{Median evolutionary track of galaxy groups in the plane of the luminosity gap and the BGG off-set ($\Delta M_{12}- D_{off}$). The evolutionary tracks are colour coded based on the age of the halo as in fig. \ref{gap_off}.  Two filled symbols within the circles mark is are used to highlight the luminosity gap (size of the symbols) and also to indicate whether the BGG is located centrally or not. The dotted-line, dotted-dashed-line and dashed-line are the mean data points of SDSS, IS-1 and MS-Guo+11, respectively.}
	\label{gap_off_model}
\end{figure*}

All of these studies rule out the possibility that a large luminosity gap has a statistical origin.
In this study, we estimate the luminosity gap parameter within a radius of 500 kpc/h for 3 galaxy group samples; SDSS (observational), MS-Guo+11 (simulations; semi-analytic) and IS-1 (simulations; hydrodynamical). In both observation and simulations, the luminosity gap is measured within a projected distance from the most luminous galaxy in the group. 

Figure \ref{gap12} shows the distribution of the luminosity gap for the SDSS groups and the two aforementioned model predictions. The distributions show that the fossil groups, i.e. galaxy groups with a large luminosity gap, are overproduced by the IS-1 in comparison to the observations and semi-analytic model used here. For instance, the fraction of fossil galaxy groups base on the definition of luminosity gap  ($\Delta m_{12}\geq$ 1.7 mag; vertical gray dashed-line) in IS-1, MS-Guo+11 and SDSS are $\approx$ 47, 22 and 27 per cent, respectively. Moreover, galaxy groups with small magnitude gaps are under produced in contrast to observation and previous studies \citep{Smith2010}. Note that, $\Delta m_{12} = 4$ mag upper limit has been adopted base on the redshift completeness in the SDSS galaxies.

As in \citet{Smith2010} and \citet{Tavasoli2011} , we present in  Figure \ref{BF:fig} the distribution of the luminosity gap as a function of the SDSS absolute $r$-band magnitudes \citep[rest frame;][]{Stoughton2002} for the first and second ranked group galaxies in the SDSS, MS-Guo+11 and IS-1. The figure suggests that in contrast to IS-1, MS-Guo+11 and SDSS follow the same trend.

 We note here that there are several adjustments that one could make in order to make SAMs consistent with a set of observational properties of galaxies. For example although all SAMs based on the Millennium simulation \citep[e.g.][]{Croton2006,Bower2006,DeLucia2007,Guo11} well reproduce general observed characteristics of galaxies such as luminosity function and colour bi-modality, which indeed are important factors to properly recover majority of observed galaxy properties, they do not fully agree on the prediction of observable parameters such as the luminosity gap  \citep[e.g. $\Delta m_{12}$ or $\Delta m_{13}$; ][]{Dariush2010}. The source of such a discrepancy, to some extent, depends on the ways different treatments have been implemented into SAMs to account for the dynamical friction (as well as other physical processes involved) in order to predict the faith of infall galaxies in groups/clusters. Indeed the main reason that the  \cite{Guo11} model has been adopted for the purpose of current study is its robustness against the luminosity gap measurements \citep[e.g.][]{Gozaliasl2014b} 

\begin{table}
	\centering
	\caption{Median luminosity gap ($\Delta M_{12}$) and centroid offset ($D_{off}$) in different redshift bins for 2 categories of galaxy groups shown in fig. \ref{gap_off_model} (1) sky blue  and (2) red. Column 1: redshift;  Column 2,4: magnitude gap between the two most luminous galaxies in the group; Column 3,5: physical separation between the BGG and the luminosity centroid of the group, centroid offset.} 
	
	\begin{tabular}{ccccc}
		\hline
		redshit   & $\Delta M_{12}(1)$ & $D_{off}(1)$	& $\Delta M_{12}(2)$ & $D_off(2)$\\ 
		-- &  [mag]   & [Mpc]   & [Mag]      & [Mpc] \\ 
		\hline\hline
		
		0.98871&	0.88478&	0.13739&	0.86139	& 0.06097\\
		0.90546&	0.81897&	0.12588&	0.95807	& 0.05898\\
		0.8277 &	0.87772&	0.13164&	0.9606	& 0.06625\\
		0.75504&	0.83571&	0.13995&	0.91877	& 0.06247\\
		0.68711&	0.82057&	0.12339&	0.81165	& 0.05092\\
		0.62359&	0.87549&	0.12641&	0.94561	& 0.04388\\
		0.56418&	0.89932&	0.12322&	0.9324	& 0.04949\\
		0.50859&	0.90665&	0.12058&	1.03455	& 0.04188\\
		0.45658&	0.98908&	0.11449&	1.0715	& 0.03431\\
		0.4079 &	1.00568&    0.12119&	1.15871	& 0.02996\\
		0.36234&	1.07564&	0.11808&	1.24518	& 0.0296\\
		0.3197 &	1.09738&	0.11   &	1.25695	& 0.02564\\
		0.2798 &	1.11613&	0.12398&	1.38156	& 0.02494\\
		0.24247&	1.17151&	0.12331&	1.58907	& 0.02589\\
		0.20755&	1.19065&	0.14448&	1.74282	& 0.02078\\
		0.1749 &	1.24234&	0.15149&	1.97987	& 0.01951\\
		0.14438&	1.27901&	0.21507&	2.12837	& 0.0189\\
		0.11588&	1.30518&	0.40083&	2.26004	& 0.01662\\
		0.08929&	1.35794&	0.82458&	2.30102	& 0.01515\\
		0.06449&	1.3548 &	0.9089 &	2.34741	& 0.0124\\
		0.0414 &	1.39455&	0.93671&	2.41839	& 0.01003\\
		0.01993&	1.4333 &	0.95849&	2.49909	& 0.00796\\
		0.0    &	1.42636&	1.00297&	2.57957	& 0.00624\\
		\hline
	\end{tabular}
	\label{gap-off:tab}
\end{table}

\subsection{Magnitude gap vs. BGG centroid offset}

In \cite{Raouf2014}, we show that the luminosity gap and the offset between the location of the BGG and the luminosity centroid are useful indicators for the dynamical age or virialization state of galaxy groups.
Thus in Figure \ref{gap_off}, we show the correlation of the luminosity gap with the inverse of centroid offset ($1/D_{off}$) for three group catalogues of SDSS, IS-1 and MS-Guo+11. Data points are colour coded according to the halo age indicators i.e.: mass assembly history ($M_{200, z=1}/M_{200, z=0}$) in case of the simulation and the age probability map in the 3D parameter plane of $\Delta m_{12}$, centroid off-set and $M_r$(BGG) (associated to different range of group galaxy luminosities) as an indicator of the halo dynamical age in case of the SDSS data, as explained in \citet{Raouf2014}.

The gray lines in each panel shows the mean of binned data points with the standard deviation errors. Broadly, the three panels in figure \ref{gap_off} show similar trends in the $\Delta M_{12}- D_{off}$ relation. In comparison to the observations, the IS-1 appears to perform more successfully in predicting the observed distribution, however, as mentioned before it significantly over-predicts the fraction of large luminosity gap galaxy systems. A possible explanation is the highly enhanced dark matter particle mass resolution of IS-1 (by a factor of $\sim100\times$) compared to the Millennium dark matter simulation as this makes the former  more robust in handling the dynamics of baryon particles.

\subsubsection{An evolutionary track for galaxy groups}

As stated above the luminosity gap and the luminosity offset complement each other to target the most evolved galaxy groups.  To better understand it we superimpose all panels in Figure \ref{gap_off}  on top of each other. This is shown in Figure \ref{gap_off_model}  where data points represent the mean and standard deviation error-bars associated to SDSS (gray dotted-line), IS-1(gray dotted-dashed) and MS-Guo+11(gray dashed-line) catalogues. We trace back the evolution of galaxy groups located at the top-right of the Figure (red model i.e. high magnitude gap $\Delta M_{12}>2$ and small centroid offset $D_{off} < 50 kpc$)  from the present epoch at $z=0$ to the snapshot corresponding to $z=1$ in MS-Guo+11. Median of the evolutionary tracks, colour coded according to the halo dynamical age (similar to bottom panel of Figure \ref{gap_off})), are also shown in Figure \ref{gap_off_model} . Hence the red colour indicates a high probability for a group to be old. 

Similarly, we trace back the evolution of galaxy groups with large luminosity gaps ($1<\Delta M_{12}<2$) and a large centroid offset ($D_{off} > 300 kpc$) (sky blue-model) between z=0 and z=1. These type of galaxy groups also end up being originated from a population of young galaxy groups with a small luminosity gap. 
Likewise, the median of the evolutionary track of sample groups is colour coded (sky-blue to blue) based on the age of halos. As indicated by colour of the track, these galaxy groups are not entirely populated by old groups. The evolution of the magnitude gap and centroid offset against the redshift in red and sky-blue galaxy models (indices 1 and 2 respectively) are summarised in Table \ref{gap-off:tab}.

\subsection{Black hole feedback} \label{BH:sec}

In the IS-1, black holes are implemented as sink particles \citep{Bellovary2010} and thus grow in mass by accreting surrounding gas or through black hole mergers and accretion. The black hole accretion is described by a Bondi-Hoyle-Lyttleton by eq. \ref{eq:BHaccr}
\begin{equation}
	\dot M_\mathrm{BH} = \frac{4 \pi \alpha G^2 M_\mathrm{BH}^2 \rho}{(c_s^2 + v_\mathrm{BH}^2)^{3/2}}  
	\label{eq:BHaccr}
\end{equation}
where $\rho$ and $c_s$  are density and sound speed of the surrounding gas, respectively, and $v_\mathrm{BH}$ is the black hole velocity relative to the gas. Also, $\alpha$ and G are the stagnation point and the gravitational constants, respectively. In the IS-1, they use a repositioning scheme for black hole sink particles that connects them to the minimum of gravitational potential, in which case, they disregard the relative gas velocity term, $v_\mathrm{BH}$, in the accretion rate \citep[See also][]{Vogelsberger2013}.

In the IS-1, the AGN feedback regulates the star formation in galaxy formation process through thermal quasar-mode (cold-mode), thermal-mechanical radio-mode(hot-mode), and radiative mode of black hole accretion. At a high accretion rate with respect to the Eddington rate, cold-mode accretion, the black hole mass grows substantially. In contrast, the low accretion rate or radio mode, the AGN jets expand hot bubbles in the surrounding halo. The radiative AGN also known as electro-magnetic feedback impact on photo-ionisation and photo-heating rates which represent the net cooling rates for a short interval of cosmic time. Moreover, this feedback only present for each black hole which locate at the state of highest accretion around the Eddington limit \citep{Sijacki2007}.

In a recent observational study of dynamically relaxed (old) and unrelaxed (young) galaxy groups, we show that relaxed systems are less luminous in radio emissivity compared to unrelaxed galaxy groups \citep{Miraghaei2014}. In addition \cite{Suresh2015} study the central galaxies of FOF groups,  based on hydrodynamical simulation, and show that the environment of central galaxies (e.g. BGGs) is influenced by the AGN feedback. They show that the radio mode feedback which inflates large hot bubbles, heats the environment of the BGG and reduces the fraction of cold gas for star formation. Another study by \cite{Genel2014} shows that AGN radio mode feedback operates as a powerful ejecting gas in most massive halos below $z = 1$ such that halos are almost devoid of gas, in disagreement with observations. Moreover, \cite{Vogelsberger2013} show that the radio mode feedback requires more power to suppress efficient cooling in massive halos compared to previous studies.

\begin{figure}
	\includegraphics[width=0.5\textwidth]{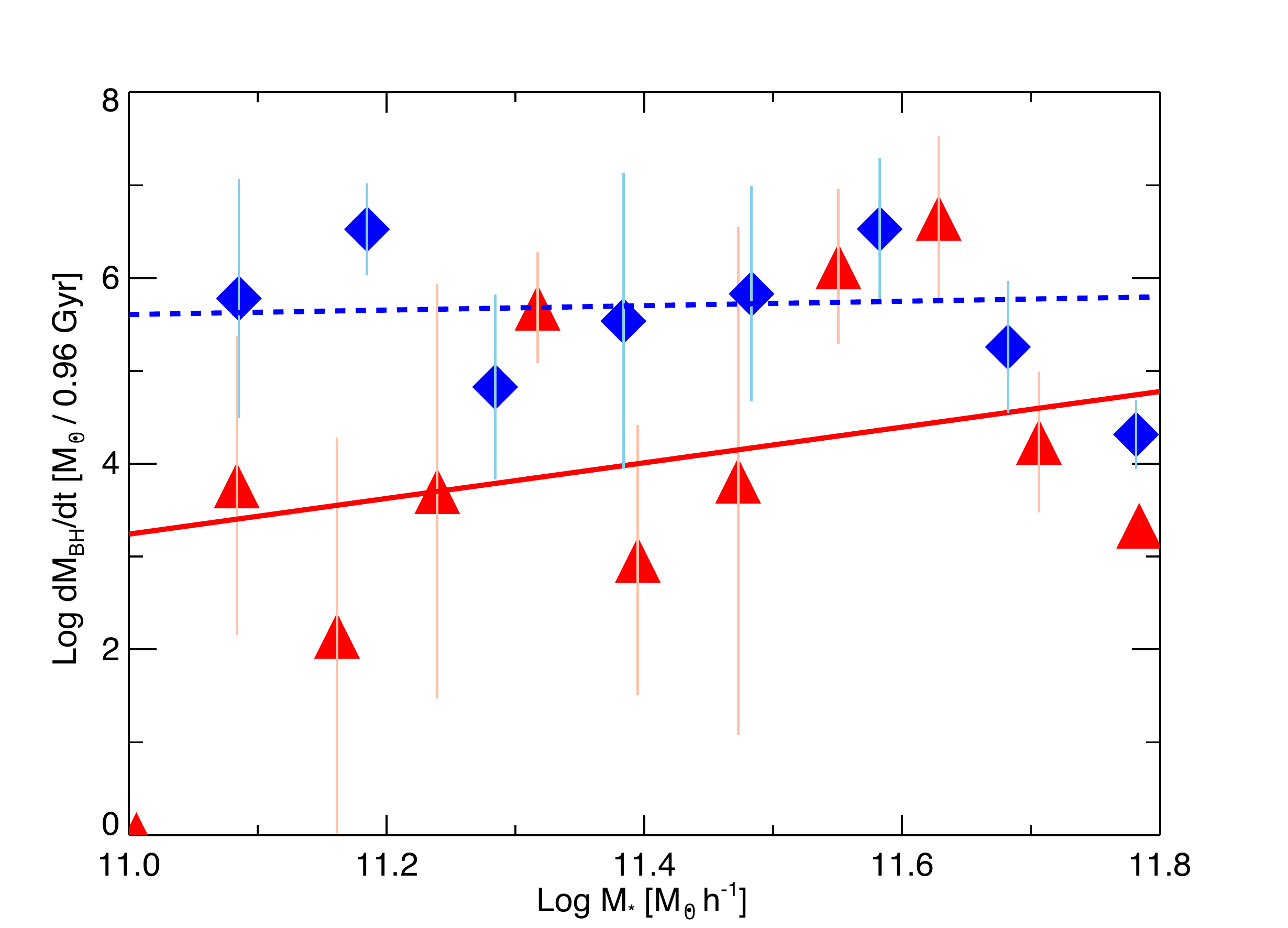}
	\includegraphics[width=0.5\textwidth]{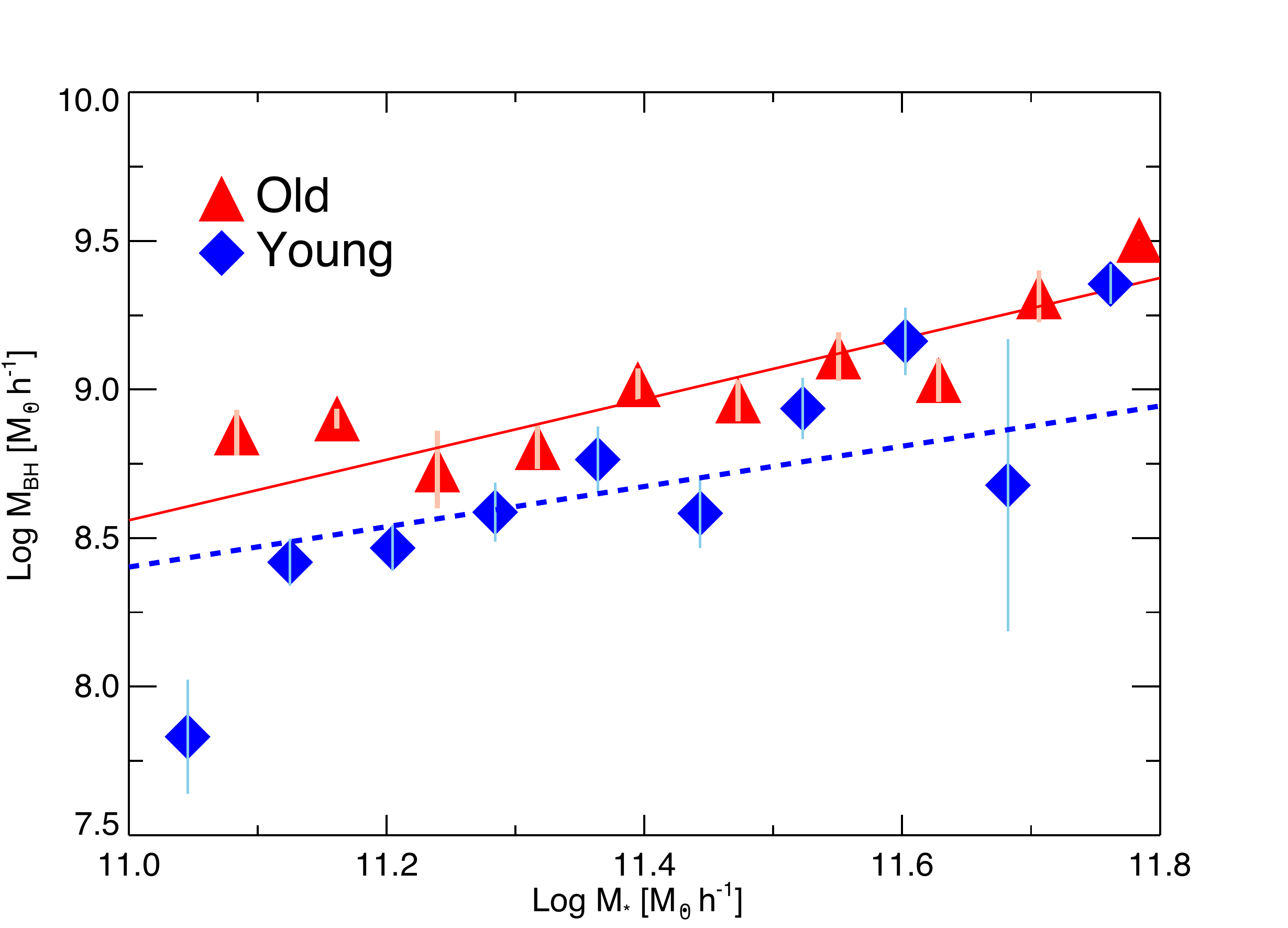}
	\includegraphics[width=0.5\textwidth]{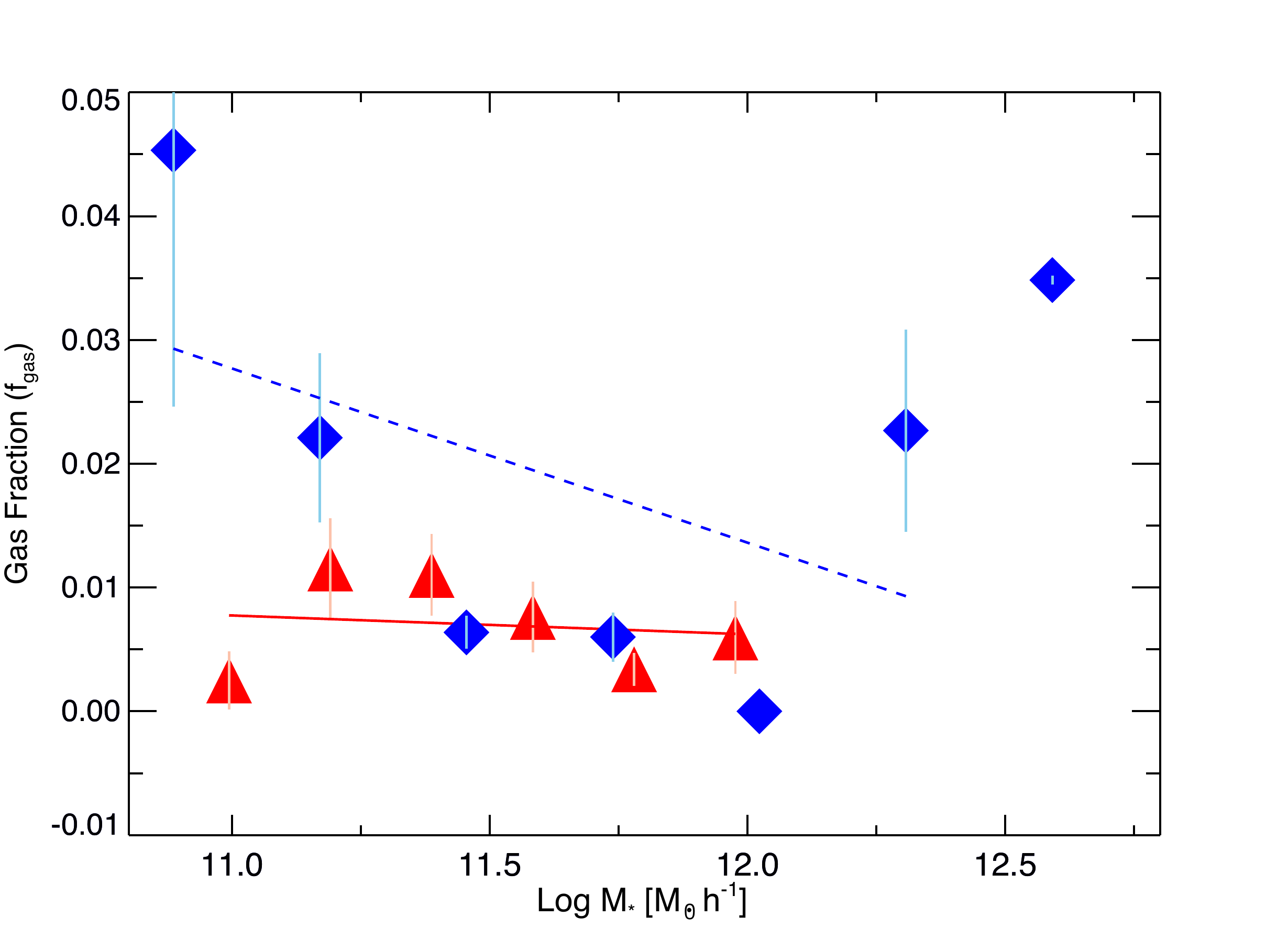}
	\caption{From top to bottom: mean values for the black hole accretion ($\dot{M}_{BH}$), black hole mass ($M_{BH}$) and gas fraction ($f_{gas}$) as a function of the BGG stellar mass for old (red) and young (blue) galaxy groups. The red line and blue dashed-line show the linear regressions to the old and young systems, respectively. The error bars in all panels are base on the standard deviation over mean, $\sigma / \sqrt{N}$. Accretion to the central black hole in the brightest group galaxies (top) is generally higher in young groups while their BGG black hole mass appears to be lower (middle). The fraction of gas in the BGG of young galaxy groups is higher than the BGG of old galaxy groups (bottom). }
	\label{BH:fig}
\end{figure}

In Figure \ref{BH:fig}, we present the distribution of the instantaneous accretion ($dM_{BH}/dt$) of all black holes (top-panel),  black hole mass (middle-panel) and gas fraction $f_{gas}$ in the sub-halo of the BGG (bottom-panel) as function of stellar mass for the BGG of old and young halos in the IS-1. As seen in Figure \ref{BH:fig}, the brightest group galaxy in the dynamically relaxed galaxy groups display a lower accretion rate compared to the brightest groups galaxies in dynamically young groups. At the same time the black hole mass in the BGGs dominating the dynamically relaxed groups is larger than in the BGGs of dynamically unrelaxed or young groups. This means that the mass assembly history of the group halos has a significant impact on the supermassive black hole of the brightest group galaxy. The BGGs dominating the dynamically old galaxy groups, seem to be very efficient in black hole growth by consuming the gas which could have been generally found with a higher density in early stages of the halo formation (as shown in bottom-panel of Figure \ref{BH:fig}). While this was argued in earlier studies of fossil groups, the Illustris provides the first direct numerical evidence.

Note that, we define the relaxed (old) and unrelaxed (young) halos base on the accumulation of $> 50$ per cent and  $< 30$ per cent of their final mass at  z $\sim$ 1, respectively. Base on this definition, we find that 33 and 29 per cent of the halos out of about 190 halos with mass over $\sim 10^{13} M_{\odot}$ within IS-1, fall in the categories of old and young groups, respectively.  The remaining halos form an intermediate population.

This is a new finding as the Illustris simulations is the first simulation, in cosmological scale, allowing us to study the growth of the supermassive black hole in the fossil dominant galaxies and also relative to giant elliptical galaxies with similar masses but in groups with a small luminosity gap. This finding makes direct connection between the dynamical state and thus the dynamical age of the halo and the growth of black hole mass. Observationally also \cite{Khosroshahi2016} show that there is a relation between the dynamical age of groups and radio luminosity of BGGs. This is following a study by \cite{Miraghaei2014} based on smaller sample. A popular argument to support the findings is based on the lack of recent on going galaxy mergers in dynamically old and fossil groups compared to their rivals, the dynamically young groups where there brightest group galaxy is expected to be surrounded by other massive galaxies \citep{Smith2010,Khosroshahi2007}. 

\subsubsection{IGM temperature}

In the IS-1, the gas temperature in each cell is obtained from the internal energy $u$ and the electron abundance $x_e$. At the first, we are estimating the mean molecular weight using eq. \ref{molec:eq}
\begin{equation}
	\mu = \frac{4}{1+3 X_H + 4 X_H x_e},
	\label{molec:eq}
\end{equation}
where $X_H$ is equal to  0.76 and present the hydrogen mass fraction. Therefore, the temperature of cells in kelvin is estimated by eq. \ref{temp:eq}

\begin{equation}
	T = (\gamma - 1) \frac{u}{K_B}(\mu m_p),
	\label{temp:eq}
\end{equation}  
where $ \gamma = 5/3 $ is the adiabatic index, and $m_p$ and $K_B$ are the proton mass and the Boltzmann constant, respectively.

\begin{figure}
	\includegraphics[width=0.5\textwidth]{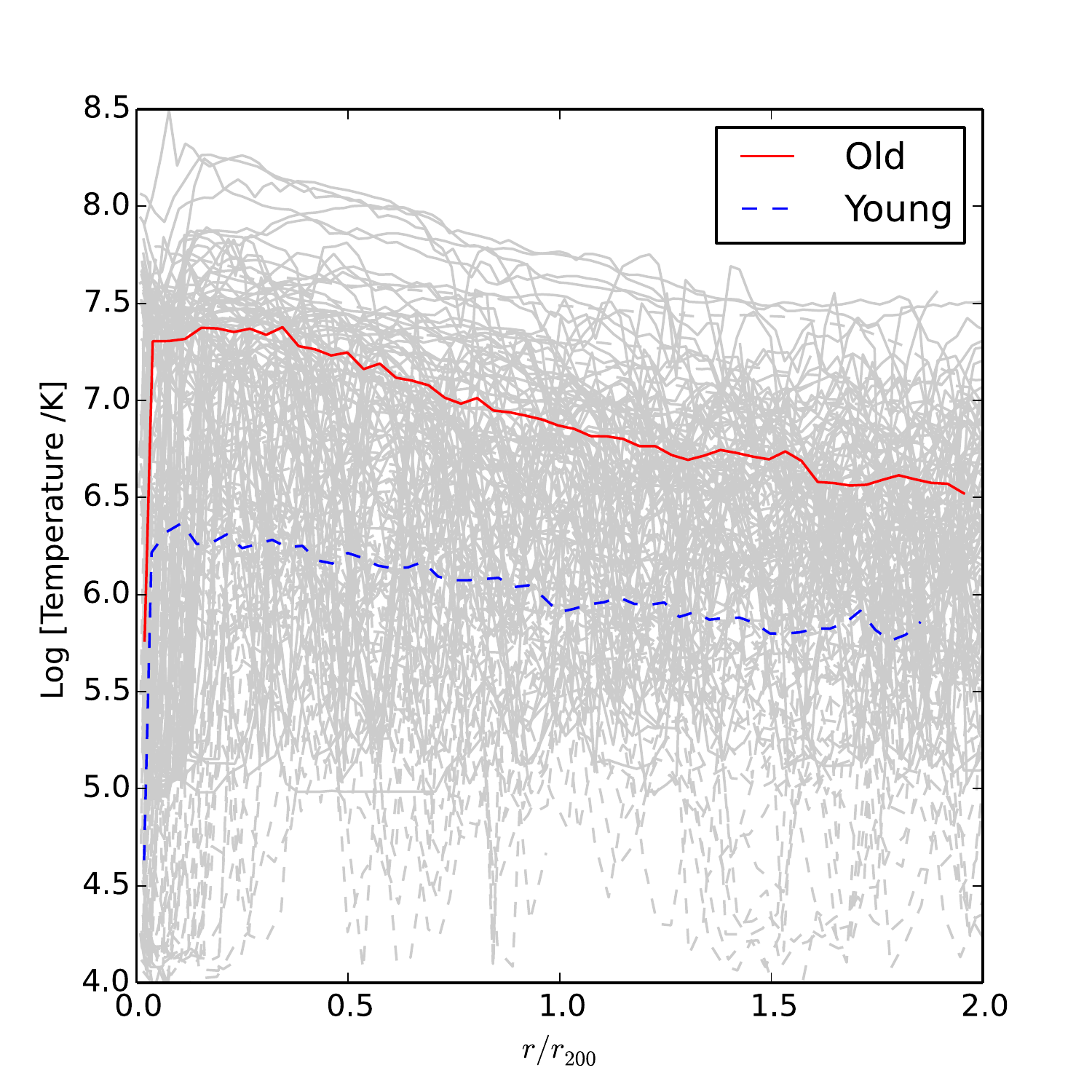}
	\includegraphics[width=0.5\textwidth]{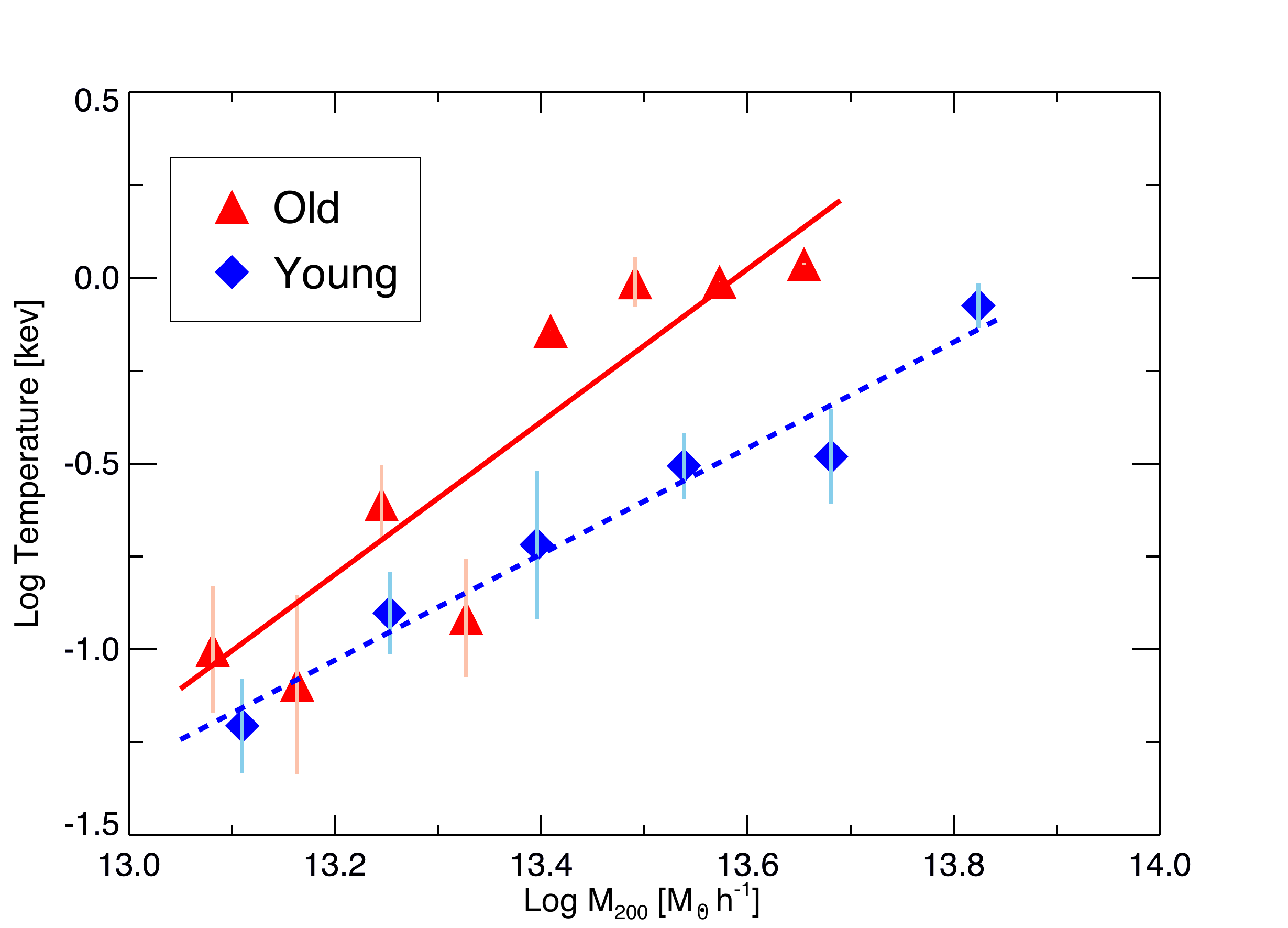}
	\caption{Top panel: IGM radial temperature profile for old (solid-line) and young (dashed-line) galaxy groups, for all the IS-1 halos with a halo mass above $10^{13} M_{\odot} h^{-1}$ at $z = 0$.  The radius is given in units of the $r_{200}$. The red-line and blue dashed-line refer to the IGM median temperature profile of old and young halos, respectively. Early formed galaxy groups have a hotted IGM in comparison to late formed galaxy groups. Bottom panel: Mean value of $M-T$ relation for old (red-triangle) and young (blue-diamond) galaxy groups. The error bars present the standard deviation over mean of bind data points in each categories. The red-line and blue-dashed line show the linear regressions to the old and young systems, respectively.}
	\label{Temp:fig}
\end{figure}

The  top-panel in Figure \ref{Temp:fig}, shows the Intra Galactic Medium (IGM) gas temperature as function of radial distance from the center in units of $r_{200}$ for dynamically old and young galaxy groups with halo masses over $10^{13} ~M_{\odot}$ at present epoch, z = 0. A comparison of the median temperature profile of old (red solid-line) and young (blue dashed-line) galaxy groups in IS-1, suggests that the IGM temperature in halos with earlier formation epoch, is systematically higher than the same in halos formed recently. In order to see if such an observed difference is due to a systematic bias in halo mass selection, we present in the lower-panel of Figure \ref{Temp:fig}  the mean value of the IGM temperature, estimated  within $r_{200}$, as a function of the halo mass, i.e. the $M-T$ scaling relation. 

From X-ray and optical observation of a sample of galaxy groups, \citet{Khosroshahi2007} show that for the same halo mass,  the IGM in fossil groups is hotter, compared to non-fossil groups.
Their sample was constructed from a small sample of fossil groups, e.g. $\Delta m_{12} \geq 2$, in which the brightest group galaxy was located at the X-ray emission pick suggesting a high degree of dynamical relaxation. Although the scale in which the IGM temperature and mass were measured in the observations, we compare to, differ from the scales shown in this study, however, we find that the results are robust and consistent with the observations. The scaling properties of relaxed and unrelaxed galaxy groups in Illustris Simulation and comparison with the observations will be presented in a separate study.

 In most SAMs, analyses of the black hole feedback are estimated in a crude way by suppression of the gas cooling provided by the radio mode of AGN \citep{Croton2006}, then act to make stellar mass function close match to the observations. The gas temperature also has to be constant and equal to the virial temperature in all processes and there are simple assumptions in it which do  not let us to use the temperature and feedback analysis for an instantaneous estimation. We try to address this issue in a separate study \citep[In prep.]{Raouf2016}  by using  the Semi Analytic Galaxy Evolution code \citep[SAGE;][]{Croton2016}  which helps to use the temperature analysis and feedback process in a more physical way which closely match the previous X-ray observations.

\section{Summary and discussion}

In this study we probe the distribution of the luminosity gap in Illustris, a new cosmological simulation, in which the properties of galaxies is dealt with hydrodynamically as opposed to a more economic, semi-analytic computation \citep{Vogelsberger2014a}. We find that galaxy groups with relatively large luminosity gaps are overproduced in Illustris with $\approx$ 47 per cent of groups having large magnitude gap ($\Delta m_{12}$) whereas the same is $\approx$ 22 per cent and  $\approx 27$ per cent in the semi-analytic model of \cite{Guo11} and the SDSS based \citet{Tempel2014} group catalogue, respectively. However, we find that the Illustris recovers the observed trend in the plane of the luminosity gap and the offset between location of brightest group galaxy and halo center of mass, as two independent indicators for the halo dynamical state. 
We show for the first time the evolutionary track of galaxy groups in the plane of  luminosity gap ($\Delta M_{12}$) vs. BGG off-set ($D_{off}$) indicating that galaxy groups with large luminosity gaps, regardless of the position  of the BGG within the group, are originated from small luminosity gap groups. However,  majority of groups with BGGs at the centre of their halos are early-formed systems.

One could argue that the higher production of large luminosity gap systems in Illustris is due to an inefficient AGN feedback. Theoretical studies suggest that AGN activity should supply enough energy to prevent gas from being accumulated in the central regions of galaxy clusters and therefore quenching the formation of stars \citep{Tabor1993,Ciotti1997,Silk1998} . According to semi-analytical models,  where the AGN feedback is paired with N-body simulations, most of the current stellar mass in the brightest cluster galaxies are assembled through dry minor mergers, following a phase of quiescent star formation influenced by feedback processes \citep{Croton2006,DeLucia2007,Bower2006,Guo11}.

Our study of the central black hole mass and the amount of mass it accreted, in old and young galaxy groups and for a given stellar mass, shows that the central black hole in fossil dominant galaxy (BGG) is noticeably more massive than a similar mass BGG in a non-fossil or a young galaxy group (e.g. Fig. \ref{BH:fig}). Furthermore, the black hole accretion in fossil dominant galaxies, on average, occurs at slower rates compared to systems with smaller luminosity gaps (i.e. non-fossil systems). This is consistent with our earlier observational findings in which fossil groups are less luminous in radio emission due to relatively less cold mode accretion \citep{Miraghaei2014}.

If galaxies in fossil groups are produced in major, possibly multiple, mergers at high redshifts, their supermassive black holes could well be more massive for their total stellar mass and as a result of no recent major merger or interaction with massive galaxies, the accretion to their central black hole could occur in a slower rate compared to galaxies which are subject to stronger interactions with more massive counterparts in groups with smaller luminosity gap \citep{Capelo2015,Lackner2014,Ellison2011,Silverman2011}. This argument is supported by a recent study in which over a dozen fossil groups were studies with a focus on their IGM properties \citep{Bharadwaj15} . 
  
The Illustris Simulation does let us for the first time to study the IGM temperature profile, suggesting that the IGM in dynamically old galaxy systems is hotter in comparison to the IGM in dynamically young/evolving halos. This is in agreement with the previous observational study of \cite{Khosroshahi2007} in which a hotter IGM was found to be associated to fossil galaxy groups, compared to non-fossil groups, with the same halo mass.

\section*{Acknowledgments}
We thank the anonymous reviewer for constructive comments which helped us to improve the manuscript. We benefited from discussions with Joseph Silk and Gary Mamon in this study.
Authors acknowledge ISEF support for this study. The Illustris project acknowledges support from many sources.
VS acknowledges support by the DFG Research Centre SFB-881 "The Milky Way System" through project A1, and by the European Research Council under ERC-StG EXAGAL-308037. GS acknowledges support from the HST grants program, number HST-AR- 12856.01-A. Support for program 12856 was provided by NASA through a grant from the Space Telescope Science Institute, which is operated by the Association of Universities for Research in Astronomy, Inc., under NASA contract NAS 5-26555. LH acknowledges support from NASA grant NNX12AC67G and NSF grant AST-1312095. DX acknowledges support from the Alexander von Humboldt Foundation. SB was supported by NSF grant AST-0907969. DN acknowledges support from XSEDE grant AST-130032, which is supported by National Science Foundation grant number OCI-1053575. Thankfully acknowledge Dylan Nelson for facilitating the access to the data. 

The Millennium  database used in this paper and the web application providing online access to them were constructed as part of the activities of the German Astrophysical Virtual Observatory.
And www.sdss3.org/. SDSS-III is managed by the Astrophysical Research Consortium for the Participating Institutions of the SDSS-III. Authors thank the anonymous referee for useful comments which helped the improvement of the text.

\label{lastpage}


\begin{thebibliography}{99}
	
	\bibitem[\protect\citeauthoryear{Aguerri et al.}{2011}]{Aguerri2011} {Aguerri}, J.~A.~L. et al, 2011,A\&A, 527A, 143
	\bibitem[\protect\citeauthoryear{Ahn et al.}{2014}]{Ahn2014}{Ahn}, C.~P. et al. , 2014, ApJS, 211, 17
	\bibitem[\protect\citeauthoryear{Barnes}{1989}]{Barnes1989} Barnes, J.E., 1989, Nature, 338, 123
	\bibitem[\protect\citeauthoryear{Bellovary al.}{2010}]{Bellovary2010}{Bellovary}, J.~M. {Governato}, F. {Quinn}, T.~R. {Wadsley}, J. {Shen}, S. {Volonteri}, M, 2010, ApJ, 721L, 148
	\bibitem[\protect\citeauthoryear{Bharadwaj et al.}{2015}]{Bharadwaj15} Bharadwaj V., Reiprich, T. H., Sanders J. S., Schellenberger, 2015, A\&A, arXiv 150904275
	\bibitem[\protect\citeauthoryear{Bode et al.}{1993}]{Bode1993}Bode, P. W., Cohn, H. N., Lugger, P. M. 1993, ApJ, 416, 17
	\bibitem[\protect\citeauthoryear{Bower et al.}{2006}]{Bower2006}{Bower}, R.~G. {Benson}, A.~J. {Malbon}, R. {Helly}, J.~C. {Frenk}, C.~S. {Baugh}, C.~M. {Cole}, S. {Lacey}, C.~G., 2006, MNRAS, 370, 645
	\bibitem[\protect\citeauthoryear{Bower et al.}{2008}]{Bower2008}{Bower}, R.~G. {McCarthy}, I.~G. {Benson}, A.~J., 2008, MNRAS, 390, 1399
	
	
	\bibitem[\protect\citeauthoryear{Capelo et al.}{2015}]{Capelo2015}{Capelo}, P.~R. {Volonteri}, M. {Dotti}, M. {Bellovary}, J.~M. {Mayer}, L. {Governato}, F., 2015, MNRAS, 447, 2123
	
	\bibitem[\protect\citeauthoryear{Ciotti \& Ostriker}{1997}]{Ciotti1997}  Ciotti L., Ostriker J. P., 1997, Astrophysical Journal Letters v.487, 487, L105
	\bibitem[\protect\citeauthoryear{Croton et al.}{2006}]{Croton2006}{Croton}, D.~J. {Springel}, V. {White}, S.~D.~M., 2006, MNRAS,365,11
	\bibitem[\protect\citeauthoryear{Croton et al.}{2006}]{Croton2016}{Croton}, D.~J.et al., 2016, ApJS, 222, 22	
	
	\bibitem[\protect\citeauthoryear{Cui et al.}{2011}]{Cui2011} Cui, Weiguang; Springel, Volker; Yang, Xiaohu; De Lucia, Gabriella; Borgani, Stefano, 2011, MNRAS, 416, 2997
	\bibitem[\protect\citeauthoryear{Davis}{1985}]{Davis1985} Davis M.,  Efstathiou G.,  Frenk C. S., White S. D. M.,  1985, ApJ, 292, 371
	\bibitem[\protect\citeauthoryear{Dariush et al.}{2007}]{Dariush2007} Dariush, A.,  Khosroshahi, H.G.,  Ponman, T.J.,  Pearce, F.,  Raychaudhury, S.,  Hartley, W.,  2007, MNRAS, 382, 433
	\bibitem[\protect\citeauthoryear{Dariush et al.}{2010}]{Dariush2010} Dariush A. A.,  Raychaudhury S.,  Ponman T. J.,  Khosroshahi H. G.,  Benson A. J.,  Bower R. G.,  Pearce F.,  2010, MNRAS, 405, 1873
	\bibitem[\protect\citeauthoryear{Deason et al.}{2013}]{Deason2013} {Deason}, A.~J. {Conroy}, C. {Wetzel}, A.~R. {Tinker}, J.~L., 2013, ApJ, 777, 154
	\bibitem[\protect\citeauthoryear{De Lucia \& Blaizot}{2007}]{DeLucia2007}{De Lucia}, G. {Blaizot}, J., 2007, MNRAS, 375, 2
	\bibitem[\protect\citeauthoryear{D{\'{\i}}az-Gim{\'e}nez et al.}{2008}]{Diaz2008} D{\'{\i}}az-Gim{\'e}nez, E.,  Muriel H., Mendes de Oliveira C., 2008, A\&A , 490, 965
	\bibitem[\protect\citeauthoryear{Dolag}{2009}]{Dolag2009}{Dolag}, K. {Borgani}, S. {Murante}, G. {Springel}, V., 2009, MNRAS,399,497
	\bibitem[\protect\citeauthoryear{D'Onghia et al.}{2005}]{DOnghia2005} D'Onghia E.,  Sommer-Larsen J.,  Romeo A. D.,  Burkert A.,  Ped- ersen K.,  Portinari L.,  Rasmussen J.,  2005, ApJ, 630, L109
	\bibitem[\protect\citeauthoryear{Ellison et al.}{2011}]{Ellison2011}Ellison S. L., Patton D. R., Mendel J. T., Scudder J. M., 2011,
	MNRAS, 418, 2043
	
	\bibitem[\protect\citeauthoryear{Genel et al.}{2014}]{Genel2014}{Genel}, S.et al.2014, MNRAS, 445, 175 
	
	\bibitem[\protect\citeauthoryear{Giodini et al.}{2010}]{Giodini2010}{Giodini}, S. et al.2010, ApJ, 714, 218
	
	\bibitem[\protect\citeauthoryear{Girardi et al.}{2014}]{Girardi2014}{Girardi}, M. {Aguerri}, J.~A.~L., et al., 2014 ,arXiv, 1403.0590
	\bibitem[\protect\citeauthoryear{Gozaliasl et al.}{2014}]{Gozaliasl2014}{Gozaliasl}, G. {Finoguenov}, A. {Khosroshahi}, H.~G., 2014, A\&A, 566A, 140
	\bibitem[\protect\citeauthoryear{Gozaliasl et al.}{2014b}]{Gozaliasl2014b}{Gozaliasl}, G. and {Khosroshahi}, H.~G. and {Dariush}, A.~A. and {Finoguenov}, A. and {Jassur}, D.~M.~Z. and {Molaeinezhad}, A., 2014, A\&A, 571A, 49
	
	\bibitem[\protect\citeauthoryear{Guo et al.}{2011}]{Guo11} Guo Q.,  White S.,  Boylan-Kolchin M., De Lucia G.,  KauffmannG.,  Lemson G.,  Li C.,  Springel V.,  Weinmann S., 2011, MNRAS, 413, 101
	\bibitem[\protect\citeauthoryear{Harrison et al.}{2012}]{Harrison2012}{Harrison}, C.~D. {Miller}, C.~J. {Richards}, J.~W. {Lloyd-Davies}, E.~J. {Hoyle}, B. {Romer}, A.~K. et al., 2012, ApJ, 752, 12
	\bibitem[\protect\citeauthoryear{Hinshaw et al.}{2013}]{Hinshaw2013}Hinshaw G. et al., 2013, ApJS, 208, 19	
	\bibitem[\protect\citeauthoryear{Jones et al.}{2003}]{Jones2003} Jones, L. R.; Ponman, T. J.; Horton, A.; Babul, A.; Ebeling, H.; Burke, D. J., 2003, MNRAS, 343, 627
	\bibitem[\protect\citeauthoryear{Khosroshahi, Jones, \& Ponman}{2004}]{Khosroshahi2004} Khosroshahi H. G.,  Jones L. R.,  Ponman T. J.,  2004,  MNRAS, 349, 1240
	\bibitem[\protect\citeauthoryear{Khosroshahi, Ponman  \& Jones}{2006}]{Khosroshahi2006} Khosroshahi H. G.,  Ponman T. J.,  Jones L. R.,  2006,  MNRAS,372, L68
	\bibitem[\protect\citeauthoryear{Khosroshahi, Ponman and Jones}{2007}]{Khosroshahi2007} Khosroshahi, H.G.,  Ponman, T.J., Jones, L.R., 2007, MNRAS, 377, 595
	\bibitem[\protect\citeauthoryear{Khosroshahi et al.}{2014}]{Khosroshahi2014}{Khosroshahi}, H.~G. et al., 2014, MNRAS, 443, 318
	\bibitem[\protect\citeauthoryear{Khosroshahi et al.}{2016}]{Khosroshahi2016},Khosroshahi H. G., Raouf M, Miraghaei H., 2016, In prep.
	\bibitem[\protect\citeauthoryear{Lackner et al.}{2014}]{Lackner2014}{Lackner}, C.~N. et al., 2014, AJ, 148, 137
	\bibitem[\protect\citeauthoryear{Nelson et al.}{2015}]{Nelson2015}{Nelson}, D. {Pillepich}, A. {Genel}, S., 2015,arXiv,150400362
	\bibitem[\protect\citeauthoryear{Milosavljevic et al.}{2006}]{Milosavljevic2006} Milosavljevic$´$,  M.,  Miller, C.J.,  Furlanetto, S.R. \& Cooray, A., 2006, ApJ, 637, L9
	\bibitem[\protect\citeauthoryear{Miraghaei et al.}{2014}]{Miraghaei2014} {Miraghaei}, H. {Khosroshahi}, H.~G. {Kl{\"o}ckner}, H.-R. {Ponman}, T.~J. {Jetha}, N.~N. {Raychaudhury}, S., 2014, MNRAS,444,651
	\bibitem[\protect\citeauthoryear{Miraghaei et al.}{2015}]{Miraghaei2015} {Miraghaei}, H. {Khosroshahi}, H.~G. and, T.~J. {Sengupta} C. {Raychaudhury}, S. {Jetha}, N.~N. and {Abbassi}, S., 2015, AJ, arXiv150908726
	\bibitem[\protect\citeauthoryear{Oklop{\v c}i{\'c} et al.}{2010}]{Oklopcic2010}{Oklop{\v c}i{\'c}}, A. et al. , 2010, ApJ, 713, 484O
	\bibitem[\protect\citeauthoryear{Proctor et al.}{2011}]{Proctor2011}{Proctor}, R.~N. {de Oliveira}, C.~M. {Dupke}, R. and 	{de Oliveira}, R.~L. {Cypriano}, E.~S. {Miller}, E.~D. {Rykoff}, E., 2011, MNRAS, 418, 2054
	\bibitem[\protect\citeauthoryear{Ponman et al.}{1994}]{Ponman1994} Ponman,  T.J.,   Allan,  D.J.,   Jones,  L.R.,  Merrifield, M.  \& MacHardy, I.M.,  1994 Nature, 369, 462
	\bibitem[\protect\citeauthoryear{Raouf et al.}{2014}]{Raouf2014}{Raouf}, M. {Khosroshahi}, H.~G. {Ponman}, T.~J. {Dariush}, A.~A. {Molaeinezhad}, A. {Tavasoli}, S., 2014, MNRAS, 442, 1578
	
	\bibitem[\protect\citeauthoryear{Raouf et al.}{2016}]{Raouf2016} Raouf M.; Shabala S. S., Croton D. J. , Khosroshahi H., Bernyk M., 2016, In prep.   
	
	\bibitem[\protect\citeauthoryear{Sales et al.}{2007}]{Sales2007} Sales, L.V.,  Navarro, J.F.,  Lambas, D.G.,  White, S.D.M. \& Croton,D.J.,  2007, MNRAS, 382, 1901
	
	\bibitem[\protect\citeauthoryear{Santos et al.}{2007}]{Santos2007}{Santos}, W.~A. {Mendes de Oliveira}, C. {Sodr{\'e}}, Jr., L., 2007, AJ, 134, 1551
	\bibitem[\protect\citeauthoryear{Sijacki et al.}{2007}]{Sijacki2007}{Sijacki}, D. {Springel}, V. {Di Matteo}, T. {Hernquist}, L.2007, MNRAS, 380, 877
	\bibitem[\protect\citeauthoryear{Silk\& Rees}{1998}]{Silk1998}  Silk \& Rees 1998      Silk J., Rees M. J., 1998, Astronomy and Astrophysics, 331,L1
	\bibitem[\protect\citeauthoryear{Silverman et al.}{2011}]{Silverman2011}{Silverman} J. D. et al., 2011, ApJ, 743, 2
	\bibitem[\protect\citeauthoryear{Smith et al.}{2010}]{Smith2010}Smith, Graham P.; et al., 2010, MNRAS, 409, 169
	\bibitem[\protect\citeauthoryear{Spergel et al.}{2003}]{Spergel2003}{Spergel}, D.~N. {Verde}, L. {Peiris}, H.~V., 2003, ApJS, 148,175
	\bibitem[\protect\citeauthoryear{Springel et al.}{2001}]{Springel2001} Springel, V.,  White, S.D.M.,  
	Tormen, G. \& Kauffmann, G.,  2001, MNRAS, 328, 726
	\bibitem[\protect\citeauthoryear{Springel et al.}{2005}]{Springel2005a}{Springel}, V. {Di Matteo}, T. {Hernquist}, L., 2005, ApJ, 620L, 79
	\bibitem[\protect\citeauthoryear{Springel et al.}{2005}]{Springel2005} {Springel}, V. , et al. 2005, Nature, 435, 629
	\bibitem[\protect\citeauthoryear{Springel et al.}{2010}]{Springel2010}{Springel}, V., 2010, MNRAS,401,791
	\bibitem[\protect\citeauthoryear{Stoughton et al.}{2002}]{Stoughton2002}{Stoughton}, C.2002,AJ,123,485
	\bibitem[\protect\citeauthoryear{Sun et al.}{2004}]{Sun2004} 	Sun, M.; Forman, W.; Vikhlinin, A.; Hornstrup, A.; Jones, C.; Murray, S. S., 2004, ApJ, 612 ,  805
	
	\bibitem[\protect\citeauthoryear{Suresh et al.}{2015}]{Suresh2015}{Suresh}, J.et al. 2015, MNRAS, 448, 895
	\bibitem[\protect\citeauthoryear{Tabor \& Binney}{1993}]{Tabor1993}  Tabor G., Binney J., 1993, MNRAS, V.263, 263, 323
	
	\bibitem[\protect\citeauthoryear{Tavasoli et al.}{2011}]{Tavasoli2011}Tavasoli, Saeed; Khosroshahi, Habib G.; Koohpaee, Ali; Rahmani, Hadi; Ghanbari, Jamshid, 2011, PASP, 123, 1
	\bibitem[\protect\citeauthoryear{Tempel et al.}{2014}]{Tempel2014} {Tempel}, E. {Tamm}, A. {Gramann}, M. {Tuvikene} et al., 2014, A\&A,566,1
	\bibitem[\protect\citeauthoryear{Ulmer et al.}{2005}]{Ulmer2005} Ulmer, M. P.; Adami, C.; Covone, G.; Durret, F.; Lima Neto, G. B.; Sabirli, K.; Holden, B.; Kron, R. G.; Romer, A. K., 2005, ApJ, 624,  124
	\bibitem[\protect\citeauthoryear{Van den Bosch et al.}{2007}]{VandenBosch2007} van den Bosch, Frank C.; et al.,  2007, MNRAS, 376, 841
	\bibitem[\protect\citeauthoryear{Voevodkin et al.}{2010}]{Voevodkin2010}{Voevodkin}, A. {Borozdin}, K. {Heitmann}, K. {Habib}, S. and 	{Vikhlinin}, A. {Mescheryakov}, A. {Hornstrup}, A. and 	{Burenin}, R., 2010, ApJ, 708, 1376
	\bibitem[\protect\citeauthoryear{Vogelsberger et al.}{2013}]{Vogelsberger2013}{Vogelsberger}, M. {Genel}, S. {Sijacki}, D. {Torrey}, P. {Springel}, V. {Hernquist}, L.2013, MNRAS, 436, 3031
	\bibitem[\protect\citeauthoryear{Vogelsberger et al.}{2014a}]{Vogelsberger2014a}Vogelsberger M. et al., 2014a, Nature, 509, 177
	\bibitem[\protect\citeauthoryear{Von Benda-Beckmann et al.}{2008}]{VonBenda-Beckmann2008} Von Benda-Beckmann A.M.,  D'Onghia E.,  et al. , 2008, MNRAS, 386, 2345
	\bibitem[\protect\citeauthoryear{Yoshioka et al.}{2004}]{Yoshioka2004} Yoshioka T.,  Furuzawa A.,  Takahashi S.,  Tawara Y.,  Sato S.,  Yamashita K.,  Kumai Y.,  2004, Advances in Space Research, 34, 2525
	
\end{thebibliography}
\end{document}